\documentclass[12pt,english]{paper}
\usepackage[T1]{fontenc}
\usepackage[latin1]{inputenc}
\usepackage{graphicx}

\makeatletter

\makeatletter

\usepackage{lineno}

\makeatletter

\usepackage{geometry}

\makeatletter

\usepackage{epsfig}

\makeatother

\makeatother

\makeatother

\usepackage{babel}
\makeatother
\begin{document}

\title{Applying Bayesian Neural Networks to Separate Neutrino Events from Backgrounds in
Reactor Neutrino Experiments}

\author{Ye Xu$^a$%
\thanks{Corresponding author, e-mail address: xuye76@nankai.edu.cn%
}, Yixiong Meng$^a$, Weiwei Xu$^a$}

\maketitle
\begin{flushleft}
$^a$Department of Physics, Nankai University, Tianjin 300071,
People's Republic of China
\par\end{flushleft}

\begin{abstract}
A toy detector has been designed to simulate central detectors in
reactor neutrino experiments in the paper. The samples of neutrino
events and three major backgrounds from the Monte-Carlo simulation
of the toy detector are generated in the signal region. The Bayesian
Neural Networks(BNN) are applied to separate neutrino events from
backgrounds in reactor neutrino experiments. As a result, the most
neutrino events and uncorrelated background events in the signal
region can be identified with BNN, and the part events each of the
fast neutron and $^{8}$He/$^{9}$Li backgrounds in the signal region
can be identified with BNN. Then, the signal to noise ratio in the
signal region is enhanced with BNN. The neutrino discrimination
increases with the increase of the neutrino rate in the training
sample. However, the background discriminations decrease with the
decrease of the background rate in the training sample.
\end{abstract}
\begin{flushleft}
\par\end{flushleft}

\section{Introduction}

The main goals of reactor neutrino experiments are to detect $\bar{\nu_{e}}\rightarrow\bar{\nu_{x}}$
oscillation and precisely measure the mixing angle of neutrino oscillation
$\theta_{13}$. The experiment is designed to detect reactor $\bar{\nu_{e}}$'s
via the inverse $\beta$-decay reaction

\begin{center}
$\bar{\nu_{e}}+p\rightarrow e^{+}+n$.
\par\end{center}

\begin{flushleft}
The signature is a delayed coincidence between $e^{+}$ and the
neutron captured signals. In the paper, only three important sources
of backgrounds are taken into account and they are the uncorrelated
background from natural radioactivity and the correlated backgrounds
from fast neutrons and $^{8}$He/$^{9}$Li. The backgrounds like the
neutrino events consist of two signals, a fast signal and a delay
signal. It is vital to separate neutrino events from backgrounds
accurately in the reactor neutrino experiments. The selection of the
neutrino events based on the cuts is a methods that the event space
is divided into two regions by a hyper-cuboid based on the cuts, and
the events inside the hyper-cuboid, called the signal region, are
regarded as neutrino events and the events outside the hyper-cuboid
are regarded as backgrounds. In fact, the backgrounds in the signal
region couldn't be rejected by the method. The Bayesian neural
networks (BNN)\cite{key-1} is an algorithm of the neural networks
trained by Bayesian statistics. It is not only a non-linear function
as neural networks, but also controls model complexity. So its
flexibility makes it possible to discover more general relationships
in data than the traditional statistical methods and its preferring
simple models make it possible to solve the over-fitting problem
better than the general neural networks\cite{key-2}. BNN has been
used to particle identification and event reconstruction in the
experiments of the high energy physics, such as
Ref.\cite{key-3,key-4,key-5}. In this paper, BNN will be applied to
discriminate the neutrino events from the background events in the
signal region in the reactor neutrino experiments.
\par\end{flushleft}

\section{The Classification with BNN\cite{key-1,key-5}}

The idea of Bayesian neural networks is to regard the process of training
a neural network as a Bayesian inference. Bayes' theorem is used to
assign a posterior density to each point, $\bar{\theta}$, in the
parameter space of the neural networks. Each point $\bar{\theta}$
denotes a neural network. In the method of the Bayesian neural network,
one performs a weighted average over all points in the parameter space
of the neural network, that is, all neural networks. The methods make
use of training data ${(x_{1},t_{1}),(x_{2},t_{2}),...,(x_{n},t_{n})}$,
where t$_{i}$ is the known label associated with data $x_{i}$. $t_{i}=0,1,...N-1$,
if there are $N$ classes in the problems of classification; $x_{i}$
has $P$ components if there are $P$ factors on which the classification
is influenced. That is the set of data $x=(x_{1},x_{2},...,x_{n})$which
corresponds to the set of target $t=(t_{1},t_{2},...,t_{n})$.The
posterior density assigned to the point $\bar{\theta}$, that is,
to a neural network, is given by Bayes' theorem

\begin{center}
\begin{equation}
p\left(\bar{\theta}\mid x,t\right)=\frac{\mathit{p\left(x,t\mid\bar{\theta}\right)p\left(\bar{\theta}\right)}}{p\left(x,t\right)}=\frac{p\left(t\mid x,\bar{\theta}\right)p\left(x\mid\bar{\theta}\right)p\left(\bar{\theta}\right)}{p\left(t\mid x\right)p\left(x\right)}=\frac{\mathit{p\left(t\mid x,\bar{\theta}\right)p\left(\bar{\theta}\right)}}{p\left(t\mid x\right)}\end{equation}

\par\end{center}

\begin{flushleft}
where data $x$ do not depend on $\bar{\theta}$, so $p\left(x\mid\theta\right)=p\left(x\right)$.
We need the likelihood $p\left(t\mid x,\bar{\theta}\right)$ and the
prior density $p\left(\bar{\theta}\right)$, in order to assign the
posterior density $p\left(\bar{\theta}\mid x,t\right)$to a neural
network defined by the point $\bar{\theta}$. $p\left(t\mid x\right)$
is called evidence and plays the role of a normalizing constant, so
we ignore the evidence. That is,
\par\end{flushleft}

\begin{center}
\begin{equation}
Posterior\propto Likelihood\times Prior\end{equation}

\par\end{center}

\begin{flushleft}
We consider a class of neural networks defined by the function
\par\end{flushleft}

\begin{center}
\begin{equation}
y\left(x,\bar{\theta}\right)=\frac{1}{1+exp\left[-s\left(x,\bar{\theta}\right)\right]}\end{equation}

\par\end{center}

where

\begin{center}
\begin{equation}
s\left(x,\bar{\theta}\right)=b+{\textstyle {\displaystyle \sum_{j=1}^{H}v_{j}tanh\left(a_{j}+\sum_{i=1}^{P}u_{ij}x_{i}\right)}}\end{equation}
 .
\par\end{center}

\begin{flushleft}
The neural networks have $P$ inputs, a single hidden layer of $H$
hidden nodes and a single output. In the particular Bayesian neural
networks described here, each neural network has the same structure.
The parameter $u_{ij}$ and $v_{j}$ are called the weights and $a_{j}$
and $b$ are called the biases. Both sets of parameters are generally
referred to collectively as the weights of the Bayesian neural networks,
$\bar{\theta}$. $y\left(x,\bar{\theta}\right)$ is the probability
that an event, $(x,t)$, belongs to the signal. So the likelihood
of $n$ training events is
\par\end{flushleft}

\begin{center}
\begin{equation}
p\left(t\mid x,\bar{\theta}\right)=\prod_{i=1}^{n}y^{t_{i}}(1-y)^{1-t_{i}}\end{equation}

\par\end{center}

\begin{flushleft}
where it has been assumed that the events are independent with each
other.
\par\end{flushleft}

\begin{flushleft}
We get the likelihood, meanwhile we need the prior to compute the
posterior density. But the choice of prior is not obvious. However,
experience suggests a reasonable class is the priors of Gaussian class
centered at zero, which prefers smaller rather than larger weights,
because smaller weights yield smoother fits to data . In the paper,
a Gaussian prior is specified for each weight using the Bayesian neural
networks package of Radford Neal%
\footnote{R. M. Neal, \emph{Software for Flexible Bayesian Modeling and Markov
Chain Sampling}, http://www.cs.utoronto.ca/\textasciitilde{}radford/fbm.software.html%
}. However, the variance for weights belonging to a given group(either
input-to-hidden weights($u_{ij}$), hidden -biases($a_{j}$), hidden-to-output
weights($v_{j}$) or output-biases($b$)) is chosen to be the same:
$\sigma_{u}^{2}$, $\sigma_{a}^{2}$, $\sigma_{v}^{2}$, $\sigma_{b}^{2}$,
respectively. However, since we don't know, a priori, what these variances
should be, their values are allowed to vary over a large range, while
favoring small variances. This is done by assigning each variance
a gamma prior
\par\end{flushleft}

\begin{center}
\begin{equation}
p\left(z\right)=\left(\frac{\alpha}{\mu}\right)^{\alpha}\frac{z^{\alpha-1}e^{-z\frac{\alpha}{\mu}}}{\Gamma\left(\alpha\right)}\end{equation}

\par\end{center}

\begin{flushleft}
where $z=\sigma^{-2}$, and with the mean $\mu$ and shape parameter
$\alpha$ set to some fixed plausible values. The gamma prior is referred
to as a hyperprior and the parameter of the hyperprior is called a
hyperparameter.
\par\end{flushleft}

\begin{flushleft}
Then, the posterior density, $p\left(\bar{\theta}\mid x,t\right)$,
is gotten according to Eqs. (2),(5) and the prior of Gaussian
distribution. Given an event with data $x'$, an estimate of the
probability that it belongs to the signal is given by the weighted
average
\par\end{flushleft}

\begin{center}
\begin{equation}
\bar{y}\left(x'|x,t\right)=\int
y\left(x',\bar{\theta}\right)p\left(\bar{\theta}\mid
x,t\right)d\bar{\theta}\end{equation}

\par\end{center}

\begin{flushleft}
Currently, the only way to perform the high dimensional integral in
Eq. (7) is to sample the density $p\left(\bar{\theta}\mid x,t\right)$
with the Markov Chain Marlo Carlo (MCMC) method\cite{key-1,key-6,key-7,key-8}.
In the MCMC method, one steps through the $\bar{\theta}$ parameter
space in such a way that points are visited with a probability proportional
to the posterior density, $p\left(\bar{\theta}\mid x,t\right)$. Points
where $p\left(\bar{\theta}\mid x,t\right)$ is large will be visited
more often than points where $p\left(\bar{\theta}\mid x,t\right)$
is small.
\par\end{flushleft}

\begin{flushleft}
Eq. (7) approximates the integral using the average
\par\end{flushleft}

\begin{center}
\begin{equation}
\bar{y}\left(x'\mid x,t\right)\approx\frac{1}{L}\sum_{i=1}^{L}y\left(x',\bar{\theta_{i}}\right)\end{equation}

\par\end{center}

\begin{flushleft}
where $L$ is the number of points $\bar{\theta}$ sampled from $p\left(\bar{\theta}\mid x,t\right)$.
Each point $\bar{\theta}$ corresponds to a different neural network
with the same structure. So the average is an average over neural
networks, and the probability of the data $x'$ belongs to the signal.
The average is closer to the real value of $\bar{y}\left(x'\mid x,t\right)$,
when $L$ is sufficiently large.
\par\end{flushleft}

\section{Toy Detector and Simulation\cite{key-4}}

\subsection{Toy Detector}

In the paper, a toy detector is designed to simulate central detectors
in the reactor neutrino experiments, such as Daya Bay experiment\cite{key-9}
and Double Chooz experiment\cite{key-10}, with CERN GEANT4 package\cite{key-11}.
The toy detector consists of three regions, and they are the Gd-doped
liquid scintillator(Gd-LS from now on), the normal liquid scintillator(LS
from now on) and the oil buffer, respectively. The toy detector of
cylindrical shape like the detector modules of Daya Bay experiment
and Double Chooz experiment is designed in the paper. The diameter
of the Gd-LS region is 2.4 meter, and its height is 2.6 meter. The
thickness of the LS region is 0.35 meter, and the thickness of the
oil part is 0.40 meter. In the paper, the Gd-LS and LS are the same
as the scintillator adopted by the proposal of the CHOOZ experiment\cite{key-11}.
The 8-inch photomultiplier tubes (PMT from now on) are mounted on
the inside the oil region of the detector. A total of 366 PMTs are
arranged in 8 rings of 30 PMTs on the lateral surface of the oil region,
and in 5 rings of 24, 18, 12, 6, 3 PMTs on the top and bottom caps.

\subsection{Monte-Carlo Simulation of Toy Detector}

The response of the neutrino and background events deposited in the
toy detector is simulated with GEANT4. Although the physical properties
of the scintillator and the oil (their optical attenuation length,
refractive index and so on) are wave-length dependent, only averages\cite{key-11}
(such as the optical attenuation length of Gd-LS with a uniform value
is 8 meter and the one of LS is 20 meter) are used in the detector
simulation. The program couldn't simulate the real detector response,
but this won't affect the result of the comparison between BNN and
the method based on the cuts.

\begin{flushleft}
According to the anti-neutrino interaction in the detector of the
reactor neutrino experiments\cite{key-12}, the neutrino events are
uniformly generated throughout Gd-LS region (see Fig. 1). The
uncorrelated background events are generated in such a way that the
fast signal energies are generated on the base of the energy
distribute of the natural radioactivity in the proposal of the Day
Bay experiment\cite{key-9}, the energies for the neutron events of
the single signal are regarded as the delay signal energies, the
delay times are uniformly generated from 2 $\mu$s to 100 $\mu$s and
the positions of the fast signal and the delay signal are uniformly
generated throughout GD-LS region. The fast neutron events are
uniformly generated throughout Gd-LS region and their energy are
uniformly generated from 0 MeV to 50 MeV, therein the events of two
signals are regarded as the fast neutron backgrounds. Since the
behavior of $^{8}$He/$^{9}$Li decay in the detector couldn't be
simulated by the Geant4 package, $^{8}$He/$^{9}$Li events are
generated in such a way that the fast signal energies are generated
on the base of the energy distribute of $^{8}$He/$^{9}$Li in the
proposal of the Day Bay experiment\cite{key-9}, and the other
physical quantities are from fast neutron events in the paper.
\par\end{flushleft}

\section{Event Reconstruction\cite{key-4}}

The task of the event reconstruction in the reactor neutrino experiments
is to reconstruct the energy and the vertex of a signal. The maximum
likelihood method (MLD) is a standard algorithm of the event reconstruction
in the reactor neutrino experiments. The likelihood is defined as
the joint Poisson probability of observing a measured distribution
of photoelectrons over the all PMTs for given ($E,\overrightarrow{x}$)
coordinates in the detector. The Ref.\cite{key-13} for the work of
the CHOOZ experiment shows the method of the reconstruction in detail.

In the paper, the event reconstruction with the MLD are performed
in the similar way with the CHOOZ experiment\cite{key-13}, but the
detector is different from the detector of the CHOOZ experiment, so
compared to Ref.\cite{key-13}, there are some different points in
the paper:

(1) The detector in the paper consists of three regions, so the path
length from a signal vertex to the PMTs consist of three parts, and
they are the path length in Gd-LS region, the one in LS region, and
the one in oil region, respectively.

(2) Considered that not all PMTs in the detector can receive photoelectrons
when a electron is deposited in the detector, the $\chi^{2}$ equation
is modified in the paper and different from the one in the CHOOZ experiment,
that is, $\chi^{2}=\sum_{N_{j}=0}\bar{N_{j}}+\sum_{N_{j}\neq0}(\bar{N}_{j}-N_{j}+N_{j}log(\frac{N_{j}}{\bar{N_{j}}}))$,
where $N_{j}$ is the number of photoelectrons received by the j-th
PMT and $\bar{N_{j}}$ is the expected one for the j-th PMT\cite{key-13}.

(3) $c_{E}\times N_{total}$ and the coordinates of the charge center
of gravity for the all visible photoelectrons from a signal are regarded
as the starting values for the fit parameters($E,\overrightarrow{x}$),
where $N_{total}$ is the total numbers of the visible photoelectrons
from a signal and $c_{E}$ is the proportionality constant of the
energy $E$, that is, $E=c_{E}\times N_{total}$. $c_{E}$ is obtained
through fitting $N_{total}$'s of the 1 MeV electron events, and is
$\frac{1}{235/MeV}$ in the paper.

\begin{flushleft}
The fast and delay signals of a event in the toy detector are
reconstructed using MLD, respectively.
\par\end{flushleft}

\section{Monte-Carlo Sample in Signal Region}

The selections of neutrino events are as follows:

(1) Positron energy: 1.3 MeV < $E_{e^{+}}$ < 8 MeV;

(2) Neutron energy: 6 MeV < $E_{n}$ < 10 MeV;

(3) Neutron delay: 2 $\mu$s < $\Delta$$t_{e^{+}n}$ < 100 $\mu$s;

(4) Relative positron-neutron distance: $d_{e^{+}n}$ < 100 cm.

A hyper-cuboid in the event space is defined by the selection, and
the inside is the signal region and the outside is the background
region. 39000 events of neutrino are generated in the signal region.
11000 events each of uncorrelated background, fast neutrons and $^{8}$He/$^{9}$Li
are generated in the signal region, respectively.

\section{Neutrino Discrimination with BNN in Signal Region}

The energies of the fast signal and the delay signal($E_{e^{+}}$,
$E_{n}$), the delay time of the delay signal($\Delta$$t_{e^{+}n}$)
and the distance between the fast signal and the delay signal ($d_{e^{+}n}$)
are used as inputs to all neural networks, which have the same structure.
In the paper, all the networks have the input layer of four inputs,
the single hidden layer of nine nodes and the output layer of a single
output which is just the probability that an event belongs to the
neutrino event. A Markov chain of neural networks is generated using
the Bayesian neural networks package of Radford Neal, with a training
sample consisting of the neutrino events and the backgrounds. One
thousand iterations, of twenty MCMC steps each, are used. The neural
network parameters are stored after each iteration, since the correlation
between adjacent steps is very high. That is, the points in neural
network parameter space are saved to lessen the correlation after
twenty steps here. It is also necessary to discard the initial part
of the Markov chain because the correlation between the initial point
of the chain and the points of the part is very high. The initial
three hundred iterations are discarded here. 3000 events each of the
neutrino and the three backgrounds are used to test the identification
capability of the trained BNN. In the paper, the BNNs are trained
by the different training samples, which consist of the neutrino events
and three backgrounds at different rates, since the different identification
efficiencies are obtained with those BNNs. The results of the identification
with those BNNs are listed in Tab. 1.

\section{Results and Discussion}

As Tab. 1, the neutrino discrimination increases from 82.6\% to
91.2\% with the increase of the neutrino rate from one second to
nine fourteenth in the training sample using BNN in signal region.
However, the background discriminations decrease with the decrease
of the background rate in the training sample. The uncorrelated
background discrimination decrease from 88.2\% to 73.6\% with the
decrease of its rate from one sixth to one fourteenth in the
training sample. The fast neutron background discrimination
decreases from 48.5\% to 37.6\% with the decreases its rate from one
sixth to one seventh in the training sample. The $^{8}$He/$^{9}$Li
background discrimination decreases from 51.8\% to 39.9\% with the
decrease of its rate from one sixth to one seventh in the training
sample. As a result, the most neutrino events and uncorrelated
background events in the signal region can be identified with BNN,
and the part events each of the fast neutron and $^{8}$He/$^{9}$Li
backgrounds in the signal region can be identified with BNN. The
different signal to noise ratios in signal region are obtained with
BNNs trained by the training samples consisting of neutrino events
and background events at different rates in the reactor neutrino
experiments. In a word, the signal to noise ratio in signal region
can be enhanced with BNN in the reactor neutrino experiments.

\section{Acknowledgements }

This work is supported by the National Natural Science Foundation
of China (NSFC) under the contract No. 10605014.

\newpage{}

\begin{table}

\caption{The different identification efficiencies are obtained with
the BNNs trained by the different training samples, which consist of
the neutrino and three backgrounds at different rates. The term
after $\pm$ is the statistical error of the identification
efficiencies. The 3000 events each of the uncorrelated background,
fast neutron and $^{8}$He/$^{9}$Li are regarded as the test sample.}
\begin{tabular}{|c|c|c|c|c|c|}\hline
 neutrino(events) & 36000& 30000&
24000& 24000& 36000\\ uncorrelated background(events) & 8000& 8000&
8000& 4000& 4000\\ fast neutron(events) & 8000& 8000& 8000& 8000& 8000\\
$^{8}$He/$^{9}$Li(events) & 8000& 8000& 8000& 8000& 8000\\\hline
neutrino eff.(\%)& 89.5$\pm$0.56& 86.7$\pm$0.62& 82.6$\pm$0.69&
85.0$\pm$0.65& 91.2$\pm$0.52\\ uncorrelated background eff.(\%)&
81.8$\pm$0.70& 84.6$\pm$0.66& 88.2$\pm$0.59& 82.6$\pm$0.69&
73.6$\pm$0.80\\ fast neutrons eff.(\%)& 39.2$\pm$0.89&
42.4$\pm$0.90& 48.5$\pm$0.91&
45.7$\pm$0.91& 37.6$\pm$0.88\\
 $^{8}$He/$^{9}$Li eff.(\%) &
40.8$\pm$0.90& 44.8$\pm$0.91& 51.8$\pm$0.91& 49.9$\pm$0.91&
39.9$\pm$0.89\\\hline
\end{tabular}
\end{table}

\begin{figure}
\includegraphics[width=16cm,height=16cm]{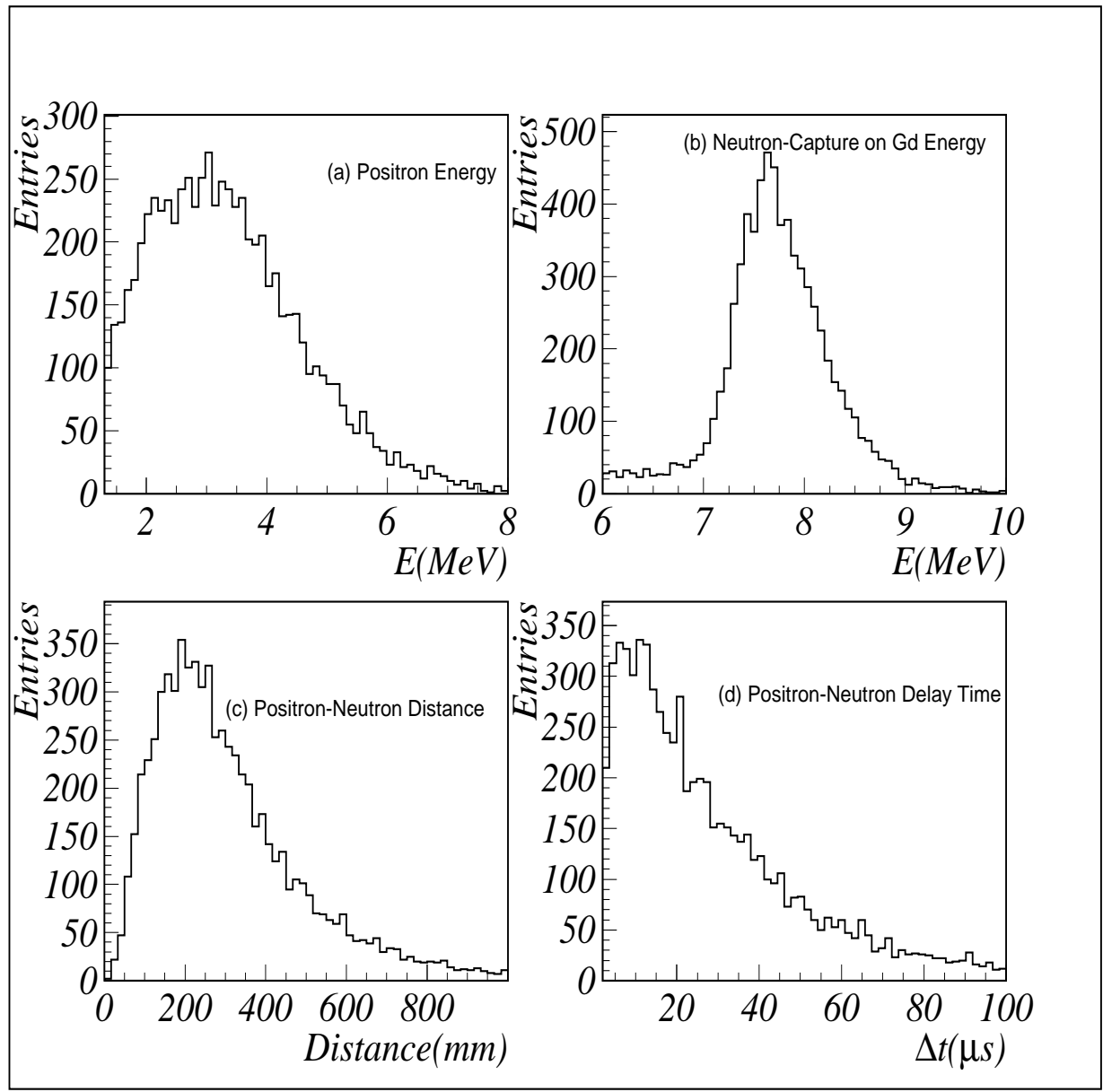}

\caption{The neutrino events for the Monte-Carlo simulation of the toy detector
are uniformly generated throughout Gd-LS region. (a) is the distribution
of the positron energy; (b) is the distribution of the energy of the
neutron captured by Gd; (c) is the distribution of the distance between
the positron and neutron positions; (d) is the distribution of the
delay time of the neutron signal.}
\end{figure}

\end{document}